\documentclass[pra,showpacs,twocolumn]{revtex4}

\usepackage{graphicx}
\usepackage{dcolumn}
\usepackage{bm}
\usepackage{color}
\usepackage{amsthm,amsmath}
\usepackage{subfig}
\usepackage{ulem} 
\usepackage{float}

\usepackage{hyperref,bbm}
\usepackage{xr}
\usepackage{xspace}

\renewcommand{\>}{\rangle}

\def\duzomniejsze{<\kern-.7mm<}
\def\duzowieksze{>\kern-.7mm>}
\def\blacksquare{\vrule height 4pt width 3pt depth2pt}

\def\textbf#1{{\bf #1}}
\def\beq{\begin{equation}}
\def\eeq{\end{equation}}
\def\be{\begin{equation}}
\def\ee{\end{equation}}
\def\ben{\begin{eqnarray}}
\def\een{\end{eqnarray}}
\def\beqa{\begin{eqnarray}}
\def\eeqa{\end{eqnarray}}
\def\eea{\end{array}}
\def\bea{\begin{array}}
\newcommand{\bei}{\begin{itemize}}
\newcommand{\eei}{\end{itemize}}
\newcommand{\bee}{\begin{enumerate}}
\newcommand{\eee}{\end{enumerate}}

\newtheorem{lemma}{Lemma}

\newtheorem{theorem}{Theorem}

\newtheorem{assmp}{Assumptions}

\def\ecal{{\cal E}}
\def\ep{\epsilon}

\begin{document}


\title{When Are Popescu-Rohrlich Boxes and Random Access Codes Equivalent?}

\author{Andrzej Grudka$^1$, Karol Horodecki$^{2}$,  Micha\l{} Horodecki$^{3}$, Waldemar K\l{}obus$^1$ and Marcin Paw\l{}owski$^{3}$}
\affiliation{$^1$Faculty of Physics, Adam Mickiewicz University, 61-614 Pozna\'n, Poland }
\affiliation{$^2$Institute of Informatics, University of Gda\'nsk, 80--952 Gda\'nsk, Poland}
\affiliation{$^3$Institute of Theoretical Physics and Astrophysics, University of Gda\'nsk, 80--952 Gda\'nsk, Poland}



\begin{abstract}
We study a problem of interconvertibility of two supra-quantum resources: one is so called PR-box,
which violates CHSH inequality up to maximal algebraic bound, and second is so called random access code (RAC).
The latter is a functionality that enables Bob (receiver) to choose one of two bits of Alice.
It has been known, that PR-box supplemented with one bit of communication can be used to simulate RAC.
We ask the converse question: to what extent RAC can simulate PR-box? To this end we introduce
racbox: a box such that supplemented with one bit of communication offers RAC. As said, PR-box can simulate racbox. The question we raise, is whether any racbox can simulate PR-box. We show that
a {\it non-signaling} racbox indeed can simulate PR-box, hence those two resources are equivalent. We also provide an
example of signalling racbox which cannot simulate PR-box. We give a resource inequality between
racboxes and PR-boxes, and show that it is saturated.

\end{abstract}

\maketitle

{\it Introduction.} Defining quantum mechanics by some information theoretic principles
have been a hot topic recently. In the seminal paper by Popescu and Rohrlich \cite{PR} 
it has been noted that the principle of no-signaling does not forbid to violate Bell inequalities
stronger than Quantum Mechanics allows. Since then much effort was devoted to answer the question,
why the systems which exhibit stronger than quantum-mechanical correlations do not exist in Nature.
The most nonlocal systems (which violate CHSH inequality maximally) are called PR-boxes.
They exhibit a variety of strange properties. One of them is that they trivialize a problem of communication complexity,
which is impossible both in quantum and classical world.
The other property is that PR-box allows for a so called random access code (RAC). Namely, suppose that
Alice has two bits and can send to Bob only one bit. Suppose further that Bob cannot communicate
to Alice. Then both in quantum and classical world, it is not possible that Bob can choose
which bit he wants to obtain and always get the right answer. However, the probability of getting it is higher if the parties have access to quantum resources.

In classical information theory RACs are basic primitives for cryptography \cite{Kilian}. In the quantum counterpart they were a basis of the first quantum protocols of Wiesner form circa 1970 (published 1983) \cite{Wiesner}. Rediscovered in \cite{Amb}, where explicit connection with the classical case was made, they were exploited for semi-device independent cryptography \cite{PB11} and randomness expansion \cite{LYWZWCGH11,M-sdirng}. They also found application in studies on foundations of quantum mechanics. RACs relation to discrete Wigner functions has been studied in \cite{Galvao} and their entanglement based version \cite{M-EARAC} in the derivation of Tsirelson bound from information-theoretic principles \cite{PawlowskiPKSWZ2009-IC}.

In \cite{PawlowskiPKSWZ2009-IC} RAC's have become a basis for Information Causality -- a principle which quantifies the success of decoding the right bit by means of mutual information. This is a new possible postulate to rule out systems
which exhibit supra-quantum correlations,
saying that the sum of mutual informations about each bit cannot exceed the number of bits
that are actually communicated.
There has been also other possible postulates (see e.g. \cite{vandam-2005,PR-ntcc,navascues-2009}). However, for a while neither of those postulates are proven to be sufficient
to ensure that a given system can be reproduced by quantum mechanics.

This development urges to further investigate supra-quantum resources
in order to understand why quantum mechanics rules them out. The two mentioned phenomena
exhibited by PR-box (trivializing communication complexity and simulating random access code)
are both of the same kind: they show that a static resource which is PR-box can simulate
some dynamical resources, RAC or possibility of computing any function with little communication.
Therefore, to have a more complete understanding of supra-quantum resources, there is a need to ask a converse question: suppose we are given some functionality, can it simulate PR-box? Thus, we ask about equivalence
between resources. The question of interconvertibility between given resources is basic for any theory of resources, e.g. entanglement theory  \cite{BBPS1996,BDSW1996,thermo-ent2002,BrandaoPlenio2007-separable-maps}, quantum communication theory \cite{AbeyesingheDHW2009-family}, or thermodynamics \cite{Beth-thermo,HO2011-thermo,BrandaoHORS20110-thermo}. Notably, following the path paved by entanglement theory, there has been done a research on interconvertion of nonsignalling boxes (see e.g. \cite{AllcockBLPSV2006-boxes-closed,BrunnerCSS2010-box-activ}).
Our present contribution goes beyond that: namely, we want to establish (in)equivalence between
nonsignalling systems (called informally {\it boxes}) on one hand and a {\it functionality} such as RAC on the other.

In this paper, we concentrate on comparison of PR-box with RAC.
As said, PR-box can simulate a racbox (i.e. an arbitrary box which supplemented with one bit of communication offers RAC). The question we raise, is whether any racbox can simulate PR-box.
We show that a non-signaling racbox indeed can simulate PR-box, hence those two resources are equivalent. We also provide an
example of signalling racbox which cannot simulate PR-box. We give a resource inequality between
racboxes and PR-boxes, and show that it is  saturated. Our paper opens a new field of study: boxes which are defined by specific tasks.

\begin{figure}
\begin{center}
\subfloat[\,]{\label{figurynka:a}\includegraphics[width=0.15\textwidth]{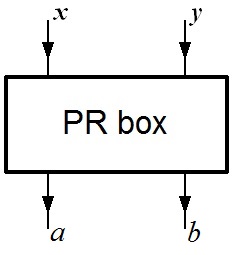}}
\qquad
\subfloat[\,]{\label{figurynka:b}\includegraphics[width=0.15\textwidth]{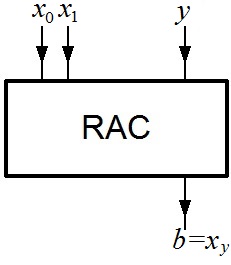}}
\\
\noindent
\subfloat[\,]{\label{figurynka:c}\includegraphics[width=0.15\textwidth]{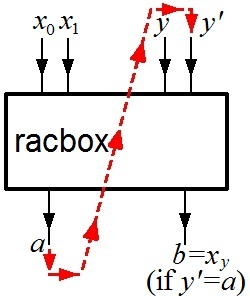}}
\qquad
\subfloat[\,]{\label{figurynka:d}\includegraphics[width=0.17\textwidth]{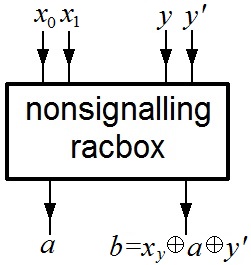}}
\end{center}
\caption{a) PR-box.
  b) RAC. c) Racbox acts as RAC, provided that the input $y'$
  is equal to $a$. Thus, in particular, if the output $a$ is sent to Bob and he inputs it
  to $y'$ (as depicted by dashed line) then $b=x_y$. d) Non-signalling racbox satisfies $b=x_y\oplus a\oplus
   y'$}
\end{figure}

{\it PR-box, random access code and racbox.}
PR-box is a bipartite system shared by two distant parties Alice and Bob.
Each of the parties can choose one of two inputs: Alice $x=0,1$ and Bob $y=0,1$.
The parties have two binary outputs $a,b$ (see Fig. \ref{figurynka:a}).
The box is defined by a family of joint probability distributions $p(ab|xy)$ which satisfy
\be
p(ab|xy)=
\left\{
\bea{l}
\frac12 \quad{\textrm{for }} a\oplus b=xy, \\
0 \quad{\textrm{else}}.
\eea
\right.
\ee
The condition
\be
\label{eq:pr-cond_1}
a\oplus b = xy,
\ee
will be called {\it PR-correlations}.


Let us now define RAC.
This is a box which has two inputs on Alice's side
(where Alice will put two bits $x_0$ and $x_1$) and no output.
On Bob's side it has an input $y$ to decide which bit
Bob wants to get $x_0$ or $x_1$, and the output $b$.
Such a box is RAC when $b=x_y$ for all possible inputs (see Fig. \ref{figurynka:b}).

It is known \cite{WW} that RAC can be simulated by PR-box assisted with one classical bit of communication. In this context one may ask whether there are other boxes of that property designed for this specific task. To this end, let us define a new type of box as in the following.

Consider a box which has in addition an output $a$ on Alice's side,
and one more input $y'$ on Bob's side (see Fig. \ref{figurynka:c}), and suppose that it is nonsignalling from Bob to Alice.
Such box we call {\it racbox}, when the following holds: if $a=y'$
then it acts as RAC on the rest of outputs/inputs, i.e.
$b=x_y$. When $a\not=y'$, we do not put any restrictions.
Racbox is thus designed in such a way, that when supplemented  with a bit of communication,
offers RAC.

PR-box is {\it non-signalling}. It means that for any choice of Bob's setting the probability distribution of his output does not depend on Alice's input and vice versa. However, in the case of racbox there is a freedom of defining the probability distribution associated with it as long as it can be turned into RAC. This makes it possible to have both signalling and nonsignalling racboxes (where signalling can be possible only from Alice to Bob).

\begin{figure}
  \begin{center}
\subfloat[\,]{\label{figurynka:2a}\includegraphics[width=0.15\textwidth]{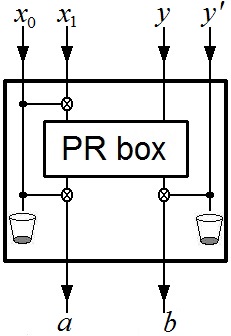}}
\qquad
\subfloat[\,]{\label{figurynka:2b}\includegraphics[width=0.15\textwidth]{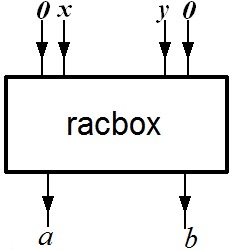}}
\end{center}
  \caption{a) Simulation of a non-signalling racbox with PR-box. b) Simulation of PR-box with nonsignalling racbox.
  We set inputs as $x_0=0,x_1=x,y'=0$, while leaving $y$ and the outputs $a$ and $b$ unchanged. This simulation precisely cancels the actions of C-NOTs in the previous one, so that we get PR-box again.}
\end{figure}

It is possible to simulate a non-signalling racbox with PR-box as illustrated in Fig. \ref{figurynka:2a}.

Now we may ask a converse question: {\it can we simulate PR-box using a racbox?}
If the answer is true, then the two resources are strictly equivalent. As we shall see, PR-box can be simulated by a {\it nonsignalling} racbox. However, we shall further present a {\it signalling} racbox which cannot simulate PR-box. Furthermore, we will derive a general resource inequality for {\it all} racboxes, and show that the signaling racbox saturates it, thus proving that the inequality is tight which reflects the fact that the signalling racbox can be considered as a weaker resource than a nonsignalling one.
Thus, \textit{all nonsignalling boxes that can perform RAC if supplemented with 1 bit of communication are equivalent to PR-box, whereas if we allow signalling there are boxes that still perform this functionality, but cannot simulate PR-box.}



{\it PR-box is equivalent to non-signaling racbox}.
Firstly let us characterize nonsignalling racboxes by the following lemma (for the proof see Appendix I \ref{seclem:nonsig-racbox}):
\begin{lemma}
A nonsignalling racbox for $a\not=y'$ operates as anti-RAC, i.e., it satisfies
\be
\label{eq:rac_antirac}
b=x_{y}\oplus a \oplus y'.
\ee
\label{lem:nonsig-racbox}
\end{lemma}

Below we will show that nonsignalling racbox can simulate PR box (see Fig. \ref{figurynka:2b}).
Namely, Alice inputs $x_0=0$, while Bob $y'=0$.
This choice is actually very natural, if one looks at the converse protocol -- of simulating racbox
with a PR-box in Fig. \ref{figurynka:2a}. The chosen fixed inputs regain the original PR-box,
i.e., they cancel the action of C-NOT gates.
Thus, in our present simulation the PR-box conditions \eqref{eq:pr-cond_1} read as
\be
\label{eq:pr-cond_2}
a\oplus b= x_1 y.
\ee
Assuming that \eqref{eq:rac_antirac} holds we proceed to show the equivalence between PR-box and nonsignalling racbox.
The PR-box condition \eqref{eq:pr-cond_2} then reads as $a\oplus x_y\oplus a \oplus y'=x_1 y$.
Recalling that in our simulation $y'=0$ we obtain a relation
\be
x_y=x_1 y,
\ee
which, since in the simulation we set also $x_0=0$,
holds for arbitrary $x_1$ and $y$ (indeed, for $y=1$ we have $x_1=x_1$ and for $y=0$
we have $x_0=0$). Therefore our simulation gives indeed a PR-box.

{\it Resource inequality between PR-box and racbox.}
We show that the following inequality holds for {\it any} racboxes:
\be
\label{eq:resource-ineq}
\textrm{racbox} + 1 \textrm{c-bit} + 1\textrm{sr-bit} \geq \textrm{PR} + \ecal,
\ee
which means that having access to any racbox (signalling or nonsignalling), one bit of communication (c-bit) and one shared random bit (sr-bit) we can simulate PR-box and additionally obtain erasure channel ($\ecal$) with probability of erasure $\ep=p(y=1)$,
where $p(y=1)$ is the probability that Bob will choose input $y=1$.

We shall prove inequality
\be
\textrm{RAC}+1 \textrm{sr-bit}\geq \textrm{PR}+1\ecal,
\label{eq:resource-ineq_2}
\ee
which implies \eqref{eq:resource-ineq}, since by definition
racbox plus 1 bit of communication offers RAC.

Let us note that to reproduce PR-correlations \eqref{eq:pr-cond_1} in case when
$y=0$ one can use just shared randomness, since the condition says that Alice and Bob's input are the same.
Thus, RAC is not used up and can be utilized to communicate the bit $x_0$.
When $y=1$, Bob will need to use RAC to reproduce PR-correlations
and in this case no communication will be performed.

\begin{figure}
  \includegraphics[width=0.18\textwidth]{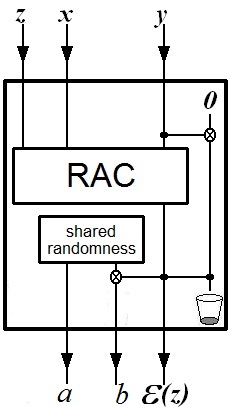}
  \caption{A protocol for achieving resource inequality \eqref{eq:resource-ineq_2}.
  The bit to be transmitted is denoted by $z$.
  $\ecal$ is erasure channel: with probability $\ep=p(y=1)$
  the message is lost, whereas with probability $1-\ep$ the message is delivered intact.
  The receiver knows which is the case. The inputs $x,y$ and the outputs $a,b$
  satisfy \eqref{eq:pr-cond_1}.}
  \label{fig:resource-ineq}
\end{figure}

Let us present the protocol which does the job (see Fig. \ref{fig:resource-ineq}).
We denote by $z$ the bit to be sent.
Alice puts $z$ to input $x_0$ and $x$ to
input $x_1$, while Bob leaves $y$ unchanged.
Regarding outputs, Alice and Bob use a shared random bit.
When $y=0$ Bob uses the random bit without any other action and, as said above,
the PR-correlations are obtained in this case.
When $y=1$ Bob performs a C-NOT on his output $b$ and the shared random bit with $b$ being the control bit and shared random bit being the target bit.
Let us see that again the PR-correlations are reproduced.
To this end, for $y=1$ we need to have correlations when $x=0$ and anti-correlations
when $x=1$.
From definition of RAC, when $y=1$, we have $b=x_1=x$.
Hence, when $x=0$, the shared random bit is not flipped, and
Alice and Bob have correlations, whereas for $x=0$ the bit is flipped,
and they have anti-correlations, as it should be.
Thus, the protocol perfectly simulates PR-box.

Let us now check how good it is regarding communication.
When $y=0$, Bob's output $b$ is equal to $x_0=z$, hence the message was perfectly transmitted,
whereas for  $y=1$, the output is equal to $x$, hence the message is lost.
Thus, we obtain erasure channel with probability of erasure $\epsilon=p(y=1)$.

{\it Tightness of the resource inequality.}
Notice that the resource inequality \eqref{eq:resource-ineq} is trivial for the case of a nonsignalling racbox. As we shall see, however, using a specific signalling racbox we can tighten the inequality (see Theorem \ref{thm:tightness} below).

We shall now present a nasty racbox which, even though performs its duty regarding RAC
(i.e. when supplemented with a bit of communication performs RAC), it cannot simulate
PR-box. Such racbox is defined as follows:
when $a=y'$, it operates as RAC (hence it is a legitimate racbox); however for $a\not=y'$ it produces
a random bit at output $b$, uncorrelated with anything else. It is signalling, because by inputting $y'=0,y=0$, Bob obtains with probability $3/4$ Alice's input $x_0$. (A particular implementation of such racbox is presented in Appendix I Fig. 4.) %

\begin{theorem}
Assume that $x$ and $y$ are generated uniformly at random.
Let us suppose that for the signalling racbox described above a channel $\Lambda$ satisfies the following inequality:
\be
\textrm{racbox} + 1 \textrm{c-bit} \geq \textrm{PR-correlations} + \Lambda.
\label{eq:resource_ineq3}
\ee
Then the channel can be obtained from $1/2$-erasure channel by postprocessing
\label{thm:tightness}
\end{theorem}
For the proof see Appendix I.
The theorem shows that in order to simulate PR-correlations by such signalling racbox we need in addition at least $1/2$ bit of communication.
Thus, in that particular instance the signaling racbox is in some respect weaker than a non-signaling one.


{\it Conclusions}. We have introduced a new functionality  called racbox. We proved that nonsignalling
racbox is equivalent to PR-box. We have also considered an exemplary signalling racbox, which, interestingly,
can be a weaker resource: in the cycle "racbox + channel $\to$ PR-box + channel" the
capacity of the channel drops at most by a half.
We have required that the output of
PR-box is perfect. It seems though possible to derive a quantitative tradeoff between
quality of PR-box and capacity of the channel (see Theorem \ref{thm:main} in Appendix II \ref{appendix:subsec-third-proof} for further details).
As an example, we can consider a more robust version where we do not aim to obtain a strict PR-correlations. In such a case one might expect a possible tradeoff between quality of PR-box and quality of a channel $z \to b$.

Our work opens a new area of studies as similar analysis can be performed not only for more general RACs but also for any other communication complexity task where nonlocal resources provide an advantage.

The most general, $n_d\to m_k$ RAC is a task in which Alice gets $n$ numbers from 1 to $d$ and sends one of $m$ possible messages to Bob, who has to guess a subset of $k$ numbers. For the simplest case studied here $n=d=2$ and $m=k=1$. If these numbers are larger the problem becomes much richer because of the freedom of which non-local box to compare with a particular racbox. One option is to consider relation between racbox and some number of PR-boxes. In this case $n_2\to 1_1$ RAC requires $n-1$ PR-boxes for simulation while being able to simulate only 1 PR-box. Another possibility is to define a generalization of a PR-box which is naturally implied by RAC. For such an entity resource inequalities analogous to the ones presented here hold. The results will be proved rigorously in \cite{prep}.

Linking non-local resources to RACs has proven to be a very powerful tool in the studies on foundations of quantum mechanics and quantum information processing protocols. Linking them to other tasks could be equally enlightening. One can, e.g. consider a {\it crypto-box} which gives the parties $N$ bits of secure key if augmented with $\mathcal{O}(N)$ bits of communication or a {\it cc-box} which gives answer to some communication complexity problem, i.e., allows the parties to find a value of a function when each of them has only a part of its input, again when augmented with some amount of two-way communication. Studies on these resources could help us understand the role of non-locality in information processing tasks.

\begin{acknowledgments}
We thank members of Gdansk QIT group for discussion, during which the problem was posed.
This work is supported by ERC grant QOLAPS, NCN grant 2013/08/M/ST2/00626 and FNP TEAM. Part of this work was done in National Quantum Information Centre of Gda{\'n}sk.
K.H. acknowledges also grant BMN nr 538-5300-B162-13.
\end{acknowledgments}


\section*{APPENDIX I: Racbox versus PR-box}\label{prufduzy}

Here we prove Lemma \ref{lem:nonsig-racbox} which says that
non-signalling racbox is equivalent to PR-box. We then present
a particular implementation of signalling racbox which is not equivalent to PR-box
as proved in Theorem \ref{thm:tightness}.
Then we prove the Theorem in two steps: (i) we will show that if the bit of communication is not used to
send $a$, but PR-correlations are obtained, then the channel $\Lambda$ is depolarizing channel: it outputs
$z$ or a random bit with probability $1/2$. (ii) If the bit of communication is used to send $a$, and PR-correlations
are obtained, then the obtainable channels $\Lambda$ have capacity no greater than $1/2$.

In what follows we will use tilde when necessary to discriminate the inputs ($\tilde{x}_0,\tilde{x}_1,\tilde{y},\tilde{y}'$) and outputs ($\tilde{a},\tilde{b}$) of racbox used in the simulation protocol from the inputs/outputs acquired by Alice and Bob ($x,y,a,b$) while simulating PR-box.

\subsection{Proof of Lemma \ref{lem:nonsig-racbox}}
\label{seclem:nonsig-racbox}

{\it Proof of Lemma \ref{lem:nonsig-racbox}.}
Suppose that Bob will choose an input $y'$ at random from $0$ or $1$. Due to non-signalling, $a$ must be independent
of $y'$. Thus, $p(a=y')=p(a\not=y')=\frac12$.
Let us now consider the probability that Bob's output $b$ is equal to $x_y$, i.e.
\begin{multline}
p(b=x_y|y)= p(a=y') p(b=x_y|a=y',y) \\
 + p(a\not=y') p(b=x_y|a\not=y',y)  \\
= \frac12 p(b=x_y|a=y',y) + \frac12 p(b=x_y|a\not=y',y).
\end{multline}
If we assume non-signalling this probability must be equal to $1/2$ for all values of $y$,
as this is precisely the probability of Bob's guessing  Alice's input $x_y$,
when he inputs $y$ and a random value of $y'$. Since for $a=y'$
the racbox operates as RAC, we have $p(b=x_y|a=y',y)=1$.
Thus, to avoid signalling we must have $p(b=x_y|a\not=y',y)=0$, i.e. when $a\not=y'$ Bob learns the
negation of $x_y$,
which can be written as $b=x_y\oplus a \oplus y'$. We thus obtain the
relation \eqref{eq:rac_antirac}.\blacksquare

\subsection{Example of a signalling racbox}

\begin{figure}[h!]
  \includegraphics[width=0.18\textwidth]{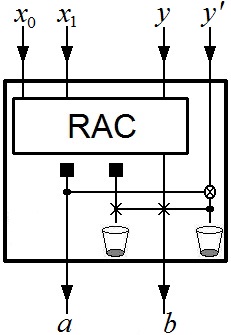}
  \caption{An example of signaling racbox which proves tightness of the inequality.
  The black squares denote generation of random bit. The gate with symbols $\times$ is
  a controlled swap gates: it swaps the two left bits if the right bit is 1.
  If $a=y'$ the racbox functions as a RAC (since then the swap is not applied, and
  Bob's output is equal to the output of RAC). When $a\not=y'$ then Bob receives a completely
  random output.}
  \label{fig:racbox-sign}
\end{figure}

In Fig. \ref{fig:racbox-sign} we present a particular implementation of the signalling racbox which cannot simulate PR-box.

\subsection{Reducing to deterministic strategies}

Here we show that in order to prove Theorem \ref{thm:tightness} it is enough to consider deterministic strategies.
\begin{lemma}
\label{lem:shared_randomness}
Consider three independent random variables $x,y,z$.
Suppose that Alice and Bob share random variable $S=(s_A, s_B)$ (where $A$ and $B$ signify Alice and Bob, respectively), which is independent of $x,y,z$.
Suppose then that Alice produces out of $x$ and $z$ two bits that she inputs to RAC as $\tilde{x}_0$ and $\tilde{x}_1$, and Bob produces $\tilde{y}$ out of $y$, and inputs it to RAC.
Consider a channel $z\to (b,s_B,y)$.

Any obtainable channel $z\to(b, s_B, y)$ is a mixture of  channels obtained by applying deterministic processing $(x,z)\to(\tilde{x}_0,\tilde{x}_1)$ and $y\to \tilde{y}$ obtained for chosen settings $(s_A,s_B)=(s^*_A,s^*_B)$.
\end{lemma}

{\it Proof.}
If Alice produces a pair $(\tilde{x}_0,\tilde{x}_1)$ from $(x,z)$ with some chosen strategy we can consider this as she applies a local channel with two-bit output $(\tilde{x}_0,\tilde{x}_1)=\Lambda_{s_A}(x,z)$, where $\Lambda_{s_A}(x,z) = \sum_i p_i \lambda^A_i(x,z)$ is a mixture of deterministic channels $\lambda^A_i(x,z)$. Similarly, Bob produces $\tilde{y}$ out of $y$ by applying a local channel $\tilde{y}=\Lambda_{s_B}(y)$, where $\Lambda_{s_B}(y) = \sum_j q_j \lambda^B_j(y)$ is a mixture of deterministic channels $\lambda^B_j(y)$.
Next, Alice and Bob inputs $\tilde{x}_0$, $\tilde{x}_1$ and $\tilde{y}$ into RAC, and Bob obtains
\begin{eqnarray}
\lefteqn{(b,s_B,y) =} \\
 &=& \Lambda_\textrm{RAC}(\tilde{x}_0,\tilde{x}_1,\tilde{y}) \nonumber \\
 &=& \displaystyle \sum_{s_A,s_B} r(s_A,s_B) \times \nonumber \\
& & \times \Lambda_\textrm{RAC} (\Lambda_{s_A}(x,z), \Lambda_{s_B}(y)) |s_B\>\<s_B| \otimes |y\>\<y| \nonumber \\
 &=& \displaystyle \sum_{s_A,s_B,i,j} r(s_A,s_B) p_i q_j \times \nonumber \\
& & \times \Lambda_\textrm{RAC} (\lambda^A_i(x,z),\lambda^B_j(y) ) |s_B\>\<s_B| \otimes |y\>\<y|, \nonumber
\end{eqnarray}
where we used Dirac notation to signify the register of chosen strategies.
\blacksquare

Moreover, if Alice produces $a$ out of $(x,\tilde{a},s_A)$ and Bob produces $b$ out of $(y,\tilde{b},s_B)$ such that they satisfy PR-correlations,
then every strategy must also reproduce PR-correlations if the mixed strategy did: if one strategy $s^*$  will fail with some probability, then the mixed strategy will also fail with some probability if $s^*$ appears in the mixed strategy.

\subsection{Proof of Theorem \ref{thm:tightness}, part (i)}

The part (i) says that if we do not input $\tilde{a}$ into $\tilde{y}'$ but require to obtain PR-correlations, the channel for $z$ is
depolarizing channel (binary symmetric channel)
with probability $1/2$ of admixing noise.

Let us denote $m$ for the one-bit message to be communicated to Bob. The goal is to obtain perfect PR-correlations $b=a\oplus xy$ in any case $m=0$ or 1.
Since the output $b$ is to be generated through the processing of RAC, then for any given $m$ its value in general depends on RAC's settings on Bob's side: $b = b (y,\tilde{y},\tilde{b})$. Now, for any fixed $m$ (let us assume $m=m_0$) there are two options: either $\tilde{a}=\tilde{y}'$ or $\tilde{a}\neq \tilde{y}'$. In the first case PR-correlations are obtained by processing a perfect RAC. However, in the case $\tilde{a} \neq \tilde{y}'$ the signalling racbox merely offers $\tilde{b}$ (and hence also $b$) which does not depend on the work of RAC, hence $b$ can be obtained solely from the processing of $y$: $b = b (y)$. Since we want to obtain perfect PR-correlations, $b$ must fulfill the conditions $b (y=0)=a$ and $b (y=1)=a \oplus x$.  Then however, by adding $b (y=0)$ and $b (y=1)$ Bob can compute $x$. We therefore obtain, that in the case $\tilde{a} \neq \tilde{y}'$, the value of $x$ must be known to Bob \cite{PawlowskiKPSB2010-PRconditions}.

We have thus proved so far, that given $m=m_0$, Bob must know either $x$ or $a$ (or both), i.e.
either $p_g(x|m=m_0)=1$ or $p_g(\tilde{a}|m=m_0)=1$ (or both), respectively, where $p_g$ denotes Bob's guessing probability.
Without loss of generality we can assume that both values of $m$ occur with nonzero probability (otherwise the channel
is not needed at all, and PR-correlations cannot be obtained, since $\tilde{a}\neq \tilde{y}'$ occurs with probability $1/2$).
Therefore, given two possible values of $m=0,1$, Bob's simplest strategy (guessing only one variable, $x$ or $\tilde{a}$, for given $m$) can rely on four different cases:
\bee
\item $p_g(\tilde{a}|m=0)=1$ and $p_g(\tilde{a}|m=1)=1$,
\item $p_g(x|m=0)=1$ and $p_g(x|m=1)=1$,
\item $p_g(\tilde{a}|m=0)=1$ and $p_g(x|m=1)=1$,
\item $p_g(x|m=0)=1$ and $p_g(\tilde{a}|m=1)=1$.
\eee

In the first case Bob makes a perfect guess of $\tilde{a}$ irrespectively of the value of $m$, in which case the one-bit message to be communicated must have been used to convey $\tilde{a}$, which enables a proper work of RAC, but also the inability to know $x$.\\
In the second case Bob makes a perfect guess of $x$ irrespectively of $m$, in which case the message was used to convey $x$, but the inability to perfectly guess the value $\tilde{a}$ affects the work of RAC.\\
In the third case (equivalently for the fourth case) Bob, depending on the value of $m$, makes a perfect guess either of $\tilde{a}$ ($p_g(\tilde{a}|m=0)=1$) or of $x$ ($p_g(x|m=1)=1$), respectively. We will see by the following example (it suffices to consider only one particular since other are analogous), however, that for the third case there cannot exist such joint probability distribution $p(\tilde{a},x,m)$ that fulfills those conditions: suppose that we want to make a perfect guess such that, e.g., $\tilde{a}=1$ given $m=0$ and $x=1$ given $m=1$. We see that the probability $p(\tilde{a}=0,x=0)$ must be 0, because each value of $m$ simply reveals the value 1 for at least one variable $\tilde{a}$ or $x$. But since $p(\tilde{a}=0,x=0)=0$, the reduced probability distribution $p(\tilde{a},x)$ is no longer randomly distributed, as it ought to be, because the box works such that $\tilde{a}$ and $x$ are generated independently at random.

From the only two possible cases we see that in the first case the inability to know $x$ forbids the proper work of PR-box, whereas in the second we get PR-box and additionally depolarizing channel with probability $1/2$.
Therefore, if we require that PR-correlations are obtained, the output $\tilde{a}$ must be sent and input to $\tilde{y}'$,
or the channel is depolarizing one.

\subsection{Proof of Theorem \ref{thm:tightness}, part (ii)}

Since the channel is used to send $\tilde{a}$, the racbox acts as RAC,
and therefore in this case Alice and Bob resource is solely  RAC plus shared randomness.
In Lemma \ref{lem:shared_randomness}
we showed, that shared randomness is not useful,
hence one should consider deterministic strategies for Alice and Bob.


There are two cases:
(a) $\tilde{y}$ does not depend on $y$;
(b) $\tilde{y}$ depends on $y$.
In first case, this means that $\tilde{y}=const$,
hence instead of RAC we have just a binary channel. However, if we have to obtain PR-correlations we would then need to know $x$ with certainty. On the other hand we cannot transmit $z$ through the binary channel at the same time, therefore the channel $\Lambda$ must have zero capacity.
Regarding (b) it is enough to consider $\tilde{y}=y$.
We use the following lemma:
\begin{lemma}
Suppose that Alice receives $x$ and Bob $y$ from the referee
with some a priori distribution $p(x,y)$.
Moreover, Bob receives $b$, that may be correlated with both $x$ and $y$.
They do not have any other resources. Then, in order to simulate PR-correlations \eqref{eq:pr-cond_1}
it must be that for any value of $b$, $p(xy|b)$ vanishes for some pair $xy$.
\label{lem:xyB}
\end{lemma}

{\textit Proof.}
W.l.o.g. we can assume that $b=0$. Suppose that all four possibilities for $(x,y)$
occur with nonzero probability. Then,  denoting $p_{min} = \min_{x,y} p(x,y)$ we have that a valid creation of
distribution $p(x,y)$ is to sample from uniform distribution $\{\frac14\}$
with probability $4p_{min}$ and from distribution $\{[p(x,y) - p_{min}]/(1-4p_{min})\}$
with probability $1-4p_{min}$. The probability of success in simulating perfectly PR-correlations
can not drop down, if Alice and Bob get to know the actual distribution of $(x,y)$.
However, in case they got uniform distribution of $(x,y)$,
and still were able to simulate PR-correlations with probability $1$,
they would violate Bell inequality, which is impossible, because in the considered
scenario, Alice and Bob do not communicate, and they initially do not share any other resource.
\blacksquare

This proof gives the structure of channels that may appear on RHS of \eqref{eq:resource-ineq}.

Now we proceed with the proof of Theorem \ref{thm:tightness}.
We use Lemma \ref{lem:xyB}. Alice's output contribute to some correlations
between $b$ and $xy$. We will now assume that Alice and Bob are able to simulate PR-box
for each output $b$. This means that for $b=0$ only three events out of
four $xyb=000,010,100,110$ will occur and for $b=1$ also only three events
out of four  $xyb=001,011,101,111$. 
Depending, on which events do not occur, we obtain three cases:
\bei
\item for $y=0$, $b$ is deterministic function of $x$;
\item for $y=1$, $b$ is deterministic function of $x$;
\item at least one value of $x$ is deterministically transmitted to $b$ (the value may
depend on $y$);
\eei
(there is also an irrelevant case for which we do not obtain all four possibilities for choosing $xy$).

\begin{table}
\begin{tabular}{ccp{1.5cm}p{1.5cm}p{1.5cm}p{1.5cm}}
\hline \hline
 $xyb$ & $xby$ & channel $x\to b$ & encoding of $x$ & encoding of $z$ & channel $z\to b$ \\ \hline
$\bea{c} 000\\ 100\\ \text{\sout{010}}\\ 110\\ \eea$      &
$\bea{c} 000\\ 100 \\ 110\\ \eea$                         &
$\bea{c} 0\to 0\\ \,\, \nearrow \,\, \\ 1\to1\\ \eea$     &
$\bea{c} x=0\\ \,\, \Downarrow \,\, \\ \tilde{x}_0=0 \\ \eea$     &
$\bea{c} x=1\\ \,\, \Downarrow \,\, \\ \tilde{x}_0=z \\ \eea$     &
$z \bea{l}\nearrow z\\ \searrow\textrm{noise}\eea $   \vspace{2mm}\\
$\bea{c} \text{\sout{001}}\\ 101\\ 011\\ 111\\ \eea$      &
$\bea{c} 101\\ 011 \\ 111\\ \eea$                         &
$\bea{c} 0\, \phantom{\to}\,  0\\ \nearrow\hspace{-3.5mm}\searrow\\ 1\to1\\ \eea$ &
$\bea{c} x=0\\ \,\, \Downarrow \,\, \\ \tilde{x}_1=1 \\ \eea$     &
$\bea{c} x=1\\ \,\, \Downarrow \,\, \\ \tilde{x}_1=z \\ \eea$     &
$z \bea{l}\nearrow z\\ \searrow\textrm{noise}\eea $\\
\hline \hline
\end{tabular}
\caption{Exemplary implication of possibility of simulating PR-correlations by RAC.}
\label{table:Bxy}
\end{table}

One finds that we can restrict to the following cases:
\begin{enumerate}
\item  for $y=0$, $b=x$;
\item  for $y=0$, $x=0$ implies $b=0$, and for $y=1$, $x=0$ implies $b=1$;
\item  for $y=0$, $x=0$ implies $b=0$, and for $y=1$, $x=1$ implies $b=1$;
\end{enumerate}
and from these three representatives we can obtain all the others by performing an appropriate bit-flip on $x$, $y$ or $b$.

In the first case we have $b=\tilde{x}_0$, hence the strategy is to put $x$
into $\tilde{x}_0$. This reduces to the protocol of Fig. \ref{fig:resource-ineq},
which implies erasure channel for $z$ with probability of erasure equal to $p(y=1)=1/2$.

The second case (elaborated schematically in Table \ref{table:Bxy}) reduces to the following protocol.
When $x=0$, then we put $0$ to $\tilde{x}_0$ and $1$ to $\tilde{x}_1$. Otherwise we put $\tilde{x}_0=\tilde{x}_1=z$.
Again we obtain the following channel for $z$ which is amplitude damping channel:
for $y=0$ the channel is: $0\to 0$ with certainty and $1 \to 0,1$ with
probability equal to $p(x=0)=1/2$. Similarly for $y=1$ where the channel is: $1\to 1$ with certainty and $0 \to 0,1$ with probability equal to $p(x=0)=1/2$.

Finally, the third case imposes the following protocol.
When $x=0$ we put it to $\tilde{x}_0$ and $z$ to $\tilde{x}_1$, and when $x=1$ we put it to $\tilde{x}_1$ and $z$ to $\tilde{x}_0$.
We obtain the following channel for $z$ which is amplitude damping channel:
for $y=0$ the channel is: $0\to 0$ with certainty and $1 \to 0,1$ with
probability equal to $p(xy=00)+p(xy=11)=1/2$. Similarly for $y=1$ where the channel is: $1\to 1$ with certainty and $0 \to 0,1$ with probability equal to $p(xy=00)+p(xy=11)=1/2$.

Given above cases we obtain basically two combinations of channels $x\to b$ and $z\to b$ while simulating PR-box, which are gathered in Table \ref{table:xz}.
\begin{table}[H]
\begin{center}
\begin{tabular}{lcc}
\hline\hline
Case No.& channel $x\to b$  & channel $z\to b$  \\ \hline
1.     & erasure & erasure \\
2.     & amplitude damping & amplitude damping \\
3.     & amplitude damping & amplitude damping \\ \hline \hline
\end{tabular}
\end{center}
\caption{Possible combinations of channels $x\to b$ and $z\to b$}
\label{table:xz}
\end{table}

\subsubsection*{Simulation of possible output channels with erasure channel}
\label{subsubsec:simluation}
Here we argue, that for $x,y$ generated uniformly at random, all kinds of amplitude damping channels can be obtained from the erasure channel of Eq. \eqref{eq:resource-ineq_2}.

To see this, consider the following erasure channel where we have two equally weighted possibilities: either bit $z$ is correctly transmitted (with a flag 0) or we obtain noise (with a flag 1), where the flags informs us which is the case. It now suffices to randomly relabel the flag 0 into 0 or 1 leaving the output $z$ intact, and randomly relabel the flag 1 and set a new output as the following: 0 for the new flag 0, and similarly 1 for the new flag 1. In such a case we obtain two equally weighted amplitude damping channels: for the flag 0 we have $z \to \{z,0\}$, and for the flag 1 we have $z \to \{z,1\}$. In order to obtain other amplitude damping channels we perform analogous procedure and we only need to establish different set of outputs for the original flag 1.

\section*{APPENDIX II: Signalling racbox versus PR-box: mutual information bound}
\label{appendix:subsec-third-proof}

We will now present another result (Theorem \ref{thm:main}), which is in a sense weaker than Theorem \ref{thm:tightness} (assumes $\tilde{a}$ to be input to $\tilde{y}'$, and does not describe possible channels), but it is more robust to possible generalizations (e.g. to obtain trade-off curves, when we do not require prefect PR correlations). Namely, we will show using information-theoretic tools,
that if the signaling racbox considered in Theorem \ref{thm:tightness} supplemented by one bit of communication is to reproduce
exactly PR box and some channel, then the mutual information of the channel must by bounded by $1/2$ (assuming that Alice's output
of the racbox will be inserted directly as Bob's second input of the racbox).

In lemmas and theorems presented here, we will consider common assumptions about scenario which we state below:

\begin{assmp}
Alice is given variables $x$ and $z$, Bob is given variable $y$, and both are given access to common variable $s$ such that $x,z,y,s$ are mutually independent.
Alice generates $a$ from $x,z$ and shared randomness $s$, and inputs $\tilde{x}_0$ and $\tilde{x}_1$ to RAC. Bob generates $\tilde{y}$ from $y$ and shared randomness $s$, and inputs it to RAC. These strategies result in shared joint probability distribution $P(x,z,y,s,\tilde{y},\tilde{b},a,b)$, where $\tilde{b}=\tilde{x}_{\tilde{y}}$ is obtained from RAC on Bob's side, and $b$ is generated out of $(\tilde{y},\tilde{b},s,y)$ by Bob.
\label{assumptions}
\end{assmp}

{\theorem Under Assumptions \ref{assumptions}, if variables $(x,y,a,b)$ perfectly reproduce PR-correlations, there holds:
\be
I(z:\tilde{b},\tilde{y},y,s) \leq {1\over 2}
\label{eq:bound-com}
\ee
where $z$ is the message that Alice sends to Bob.
\label{thm:main}
}

To prove the above Theorem we explore two ideas.
First, after \cite{PawlowskiKPSB2010-PRconditions} we rephrase in terms of entropies and correlations the fact that to simulate PR-correlations, Bob has to guess perfectly certain values given values of $y$: for $y=0$ he should guess perfectly $a$, and for $y=1$ he should guess perfectly $a \oplus x$ (see Lemma \ref{lem:max-ims}). 
Second idea sounds almost as tautology: it is impossible to send more than 1 bit through a channel with 1-bit capacity. In our case Alice would like to send both $x$ (to enable simulation of PR-correlations) and $z$, which bounds Bob's possible correlations with $z$ as stated (see Theorem \ref{thm:dependence}).


{\lemma
Under Assumptions \ref{assumptions}, if variables $(x,y,a,b)$ simulate perfectly PR-correlations, there holds:
\ben
I( \tilde{b} : a\oplus x|\tilde{y},s, y = 1) &=& H(a\oplus x|\tilde{y},s,y=1), \label{eq:max-ims1}\\
I( \tilde{b} : a| \tilde{y},s, y=0) &=& H(a|\tilde{y},s,y=0).
\een
\label{lem:max-ims}
}

{\it Proof}.
To show this, we use approach of \cite{PawlowskiKPSB2010-PRconditions}, according to which the sender creates a message $\cal X$, while the receiver upon this value tries to guess some variable $\cal Y$. Maximal probability
of correctly guessing $\cal Y$, called {\it guessed information}, reads:
\be
J({\cal X} \rightarrow {\cal Y}) = \sum_i p({\cal X}=i)\max_{j}[p({\cal Y}=j|{\cal X}=i)].
\ee
In \cite{PawlowskiKPSB2010-PRconditions} it is studied when Alice and Bob violate CHSH inequality with the help of the message $\cal X$ from one party to the other.
Using guessed information the CHSH inequality \cite{CHSH} can be rephrased as follows:
\be
{1\over 2} J({\cal X},s \rightarrow a) + {1\over 2} J{(\cal X},s \rightarrow a\oplus x) \leq {3\over 4}.
\label{eq:chsh}
\ee
Adapting this scheme to our situation, we have that  Alice and Bob are given $x$ and $y$, then Alice produces $a$ from $x$ and $s$, then inputs $\tilde{x}_0$ and $\tilde{x}_1$ to RAC, whereas Bob produces $\tilde{y}$ from $y$ and obtains some message from Alice via RAC, which is bit $\tilde{b}$. Thus, in our case ${\cal X} = \tilde{b}$, while
other variables $\tilde{y},s$ are local for Bob, however, we can w.l.o.g. treat them as a message, since Bob uses them to guess $a$ and $a\oplus x$. This leads to CHSH inequality as follows:
\be
{1\over 2} J(\tilde{b},\tilde{y},s,y=0 \rightarrow a) + {1\over 2} J(\tilde{b},\tilde{y},s,y=1 \rightarrow a\oplus x) \leq {3\over 4}.
\label{eq:chsh2}
\ee

Now, in order to reproduce PR-correlations given $y=0$, Bob should perfectly guess $a$, whereas given $y=1$ he should perfectly guess $a\oplus x$. Thus, both terms on LHS of \eqref{eq:chsh} should be equal to 1.
This implies in particular that there must be $\max_{j}[p(a=j|\tilde{b}=l,\tilde{y}=k,y=0,s=i)] = 1$. Then, for $y=0$ the values of variables $\tilde{b},\tilde{y},s$ determine uniquely the value of $a$, i.e., $H(a|\tilde{b},\tilde{y},s,y=0)=0$. In such a case $I(a:\tilde{b}|\tilde{y},s,y=0)= H(a|\tilde{y},s,y=0)$. Analogously, we obtain $I(a\oplus x:\tilde{b}|\tilde{y},s,y=1)= H(a\oplus x|\tilde{y},s,y=1)$.\blacksquare

\subsection{One cannot send more than one bit through a single-bit wire}

In this section, we prove Theorem \ref{thm:dependence} which provides the main argument in the proof of Theorem \ref{thm:main}.
Namely, it shows a tradeoff between Bob's correlations with $a$ and $a\oplus x$ (that should be high if he simulates PR correlations) and his correlations with $z$.

{\theorem\label{thm:dependence} Under assumptions \ref{assumptions}, there holds:
\begin{multline}\label{eq:dependence}
{1\over 2}I(a\oplus x : \tilde{b}|\tilde{y},s,y=1) +{1\over 2}I(a:\tilde{b}|\tilde{y},s,y=0) \\
+I(z:\tilde{b}|\tilde{y},s,y) \leq {1\over 2}I(a: a\oplus x:z|\tilde{y},s) + H(\tilde{b}|\tilde{y},s,y).
\end{multline}
}

In the proof of the above theorem, we use numerously the following fact, which captures that one cannot send reliably 2 bits through a single-bit wire, unless the bits are correlated:

{\lemma
For any random variables S,T,U,V there holds:
\begin{multline}
I(S:T|V) + I(T:U|V) \\
\leq I(S:U|V) + I(T:SU|V) \equiv I(S:T:U|V),
\end{multline}
where $I(S:T:U|V) = H(S|V) + H(T|V)+H(U|V) - H(STU|V)$.
\label{lem:transit}
}

{\it Proof}.
We first prove the above fact without conditioning. It follows directly from strong subadditivity:
\be
H(STU) + H(T) \leq H(ST) + H(SU).
\ee
Indeed, by expressing mutual information via Shannon entropies, we obtain that we need to prove:
\be
H(SU) + H(T) - H(ST) - H(TU) + H(T) \leq I(T:SU).
\label{eq:medium}
\ee
Now, by strong subadditivity LHS is bounded by
\begin{multline}
H(SU) + H(T) - \left( H(STU) + H(T) \right) + H(T)  \\
= H(SU) + H(T) - H(STU),
\end{multline}
which is RHS of (\ref{eq:medium}), proving the thesis without conditioning on $V$. We can now fix $V=v$, and the thesis will hold for conditional distribution $p(STU|V=v)$:
\begin{multline}
I(S:T|V=v) + I(T:U|V=v) \\
\leq I(S:U|V=v) + I(T:SU|V=v).
\end{multline}
The thesis is obtained after multiplying each side by $p(V=v)$, and summing over range of variable $V$.\blacksquare

{\it Proof of Theorem \ref{thm:dependence}}.

Let us first reformulate LHS of the thesis, and fix $s=i$:
\begin{multline}
{1\over 2}I(a\oplus x : \tilde{b}|\tilde{y},s=i,y=1)
+{1\over 2}I(a:\tilde{b}|\tilde{y},s=i,y=0) \\ + I(z:\tilde{b}|\tilde{y},s=i,y).
\label{eq:LHSthm}
\end{multline}
By decomposing the last term into two, which depend on the value of $y$ we obtain:
\begin{multline}
{1\over 2}[ I(a\oplus x : \tilde{b} |\tilde{y},s=i,y=1)+ I(z:\tilde{b}|\tilde{y},s=i,y=1)   \\
 + I(a:\tilde{b}|\tilde{y},s=i,y=0) + I(z:\tilde{b}|\tilde{y},s=i,y=0)].
\end{multline}
We use Lemma \ref{lem:transit} to the first and the second pair of these terms to show that the above quantity is upper bounded by
\begin{multline}
{1\over 2}[I(a\oplus x : z|\tilde{y},s=i,y=1) \\
+ I(\tilde{b}: a\oplus x,z |\tilde{y},s=i,y=1) \\
+I(\tilde{b}:a,z|\tilde{y},s=i,y=0) + I(a:z|\tilde{y},s=i,y=0)],
\label{eq:itermediate}
\end{multline}
Now, we observe that $(a\oplus x,z|s=i)$ is independent from $(y,\tilde{y}|s=i)$, hence there is $I(a\oplus x : z|\tilde{y},s=i,y=1)= I(a\oplus x: z|\tilde{y},s=i)$, and since $(a,z|s=i)$ is independent from $(y,\tilde{y}|s=i)$, there is $I(a:z|\tilde{y},s=i,y=0)= I(a|\tilde{y},s=i)$.
Multiplying both sides of these equalities by $p(s=i)$ and summing over values of $s$ we get $I(a\oplus x : z |\tilde{y},s,y=1)=I(a\oplus x: z|\tilde{y},s)$ and $I(a:z|\tilde{y},s,y=0)=I(a:z|\tilde{y},s)$. Applying the same operation to (\ref{eq:itermediate}), and using the latter equalities we obtain:
\begin{multline}
{1\over 2}[I(a\oplus x : z|\tilde{y},s)+ I(\tilde{b}: a\oplus x,z |\tilde{y},s,y=1)  \\
+I(\tilde{b}:a,z|\tilde{y},s,y=0) + I(a:z|\tilde{y},s)],
\end{multline}
so that we can use again Lemma \ref{lem:transit} to the first and last term of the above formula to obtain:
\begin{multline}
 {1\over 2}[I(a\oplus x : a|\tilde{y},s)+I(z:a,a\oplus x|\tilde{y},s)  \\
 + I(\tilde{b}: a\oplus x,z |\tilde{y},s,y=1)+ I(\tilde{b}:a,z|\tilde{y},s,y=0)].
\end{multline}
The first two terms add up exactly to $I(a:a\oplus x:z|\tilde{y},s)$, while  the last two terms are bounded by $H(\tilde{b}|\tilde{y},s,y=1)$ and $H(\tilde{b}|\tilde{y},s,y=0)$, respectively, which, because of the factor ${1\over 2}$, give rise to $H(\tilde{b}|\tilde{y},s,y)$, and the assertion follows.
\blacksquare

\subsection{Proof of Theorem \ref{thm:main}.}
We prove now the main result, which is Theorem \ref{thm:main}. To this end we first observe that in fact it is sufficient to show:
\be
I(z:\tilde{b}|y,\tilde{y},s) \leq {1\over 2}.
\label{eq:main-toprove}
\ee
Indeed, from the chain rule: $I(z:\tilde{b},\tilde{y},s,y)=I(z:y,\tilde{y},s) + I(z:\tilde{b}|\tilde{y},s,y)$, but $I(z:y,\tilde{y},s)=0$, since $I(z:y,s)=0$ by assumption, and $I(z:y,s) = I(z:y,s,\tilde{y})$ ($\tilde{y}$ emerges from $y,s$ according to Bob's strategy). Hence we get:
\be
I(z:\tilde{b},\tilde{y},s,y)=I(z:\tilde{b}|s,y,\tilde{y})\leq {1\over 2},
\ee
which is desired bound. To show (\ref{eq:main-toprove}), we use Theorem \ref{thm:dependence}, and Lemma \ref{lem:max-ims}.
From Theorem \ref{thm:dependence} we have:
\begin{multline}
{1\over 2}I(a\oplus x : \tilde{b}|\tilde{y},s,y=1) \\
+ {1\over 2}I(a:\tilde{b}|\tilde{y},s,y=0) +  I(z:\tilde{b}|\tilde{y},s,y) \\
\leq {1\over 2}I(a\oplus x : a:z|\tilde{y},s) + H(\tilde{b}|\tilde{y},s,y).
\label{eq:final-bound}
\end{multline}
As we argued, in Lemma \ref{lem:max-ims}
what follows from maximal violation of CHSH is that the first two terms of the (\ref{eq:final-bound}) are equal to ${1\over 2}H(a\oplus x|\tilde{y},s,y=1)$ and ${1\over 2}H(a|\tilde{y},s,y=0)$, respectively.
Thus, substituting this in LHS of (\ref{eq:final-bound}) and expanding the first term of its RHS, we get:
\begin{multline}\label{31}
{1\over 2}[H(a|\tilde{y},s,y=0)
 + H(a\oplus x|\tilde{y},s,y=1)] + I(z:\tilde{b}|\tilde{y},s,y) \\
  \leq {1\over 2}[H(a|\tilde{y},s)+H(a\oplus x|\tilde{y},s) + H(z|\tilde{y},s)  \\
  - H(a,a\oplus x, z|\tilde{y},s)] + H(\tilde{b}|\tilde{y},s,y).
\end{multline}

Now, because $(a|s=i)$ and $(a\oplus x|s=i)$ are independent from $(\tilde{y},y|s=i)$, we have for each $i$ that $H(a|\tilde{y},y=0,s=i)=H(a|\tilde{y},s=i)$, $H(a\oplus x|\tilde{y},y=1,s=i)=H(a\oplus x|\tilde{y},s=i)$, and because for fixed $s=i$, $z$ is independent from $\tilde{y}$, there is $H(z|\tilde{y},s=i)=H(z|s=i)$. Averaging these equalities over $p(s=i)$, we obtain that the first two terms of LHS and RHS of \eqref{31} cancel each other respectively and the inequality reads:
\begin{multline}
I(z:\tilde{b}|y,\tilde{y},s) \\
 \leq {1\over 2}[ H(z|s) - H(a,a\oplus x, z|\tilde{y},s)] + H(\tilde{b}|\tilde{y},s,y).
\end{multline}
Since $z$ is independent form $s$, $H(z|s)=H(z)=1$. Now, $H(a,a\oplus x, z|s)$ equals $H(z,a,x|s)$ as we can add $a$ to $a\oplus x$ reversibly. From the data processing inequality and the independence of $s$ from $(x,z)$, we get $H(z,a,x|s) \geq H(z,x|s) = H(z,x) = 2$, hence the first two terms are bounded from above by $-{1\over 2}$. The last term is trivially upper bounded by 1, which gives desired total upper bound ${1\over 2}$, proving \eqref{eq:main-toprove} as required.\blacksquare

\bibliographystyle{apsrev}
\bibliography{references}

\end{document}